\documentclass[12pt]{article}
\newcommand{\grouplabel}[1]{\refstepcounter{equation}\label{#1}%
\addtocounter{equation}{-1}}%
\newcounter{saveeqn}%
\newcommand{\steqnlett}{\setcounter{saveeqn}{\value{equation}}%
\stepcounter{saveeqn}\setcounter{equation}{0}%
\renewcommand{\theequation}{\arabic{saveeqn}\alph{equation}}}%
\newcommand{\eneqnlett}{\setcounter{equation}{\value{saveeqn}}%
\renewcommand{\theequation}{\arabic{equation}}}%
\newenvironment{mathletters}{\steqnlett}{\eneqnlett}%

\begin{document}

\title{Generalizing Quantum Mechanics for Quantum Gravity\footnote{This contribution to the proceedings of the Glafka Conference is an extended abstract of the author's talk there. More details can be found in the references cited at the end of the abstract expecially \cite{leshouches}.}}
\author{James B.~Hartle\\
Department of Physics, University of California,\\
Santa Barbara, CA 93106--9530}
\date{}

\maketitle

How do our ideas about quantum mechanics affect our understanding of spacetime? 
This  familiar question leads to quantum gravity. This talk addressed a complementary question: How do our ideas about spacetime affect our understanding of quantum mechanics?

Familiar non-relativistic quantum theory illustrates how quantum mechanics incorporates   assumptions about spacetime. The Schr\"odinger equation governs the evolution of the state in between measurements
\grouplabel{lawsofev}
\begin{mathletters}
 \begin{equation}
 i \hbar \frac{\partial \psi}{\partial t} = H \psi \  .
 \label{seqn}
 \end{equation}
 The state vector is ``reduced'' at the time of a measurement according to the second law of evolution:
 \begin{equation}
 \psi \rightarrow P\psi /|| P\psi|| 
 \label{sndlaw}
 \end{equation}
 \end{mathletters}%
 where $P$ is the projection on the outcome of the measurement. Both of these laws of evolution assume a fixed background spacetime. A fixed geometry of spacetime is needed to define both the $t$ in the Schr\"odinger equation and the spacelike surface on which the state vector is reduced. 
 
 But, in quantum gravity, the geometry of spacetime is not fixed. Rather geometry is a quantum variable, fluctuating and generally without a definite value. There is no fixed $t$. 
Quantum mechanics must therefore be generalized to deal with quantum spacetime. This is sometimes called the `problem of time' in quantum gravity.

Already our ideas about quantum theory have evolved as out ideas about spacetime have changed. 
 Milestones in the evolution of our concepts of space and time include: the
separate space and absolute time of Newtonian physics, Minkowski spacetime
with different times in different Lorentz frames, the curved but fixed
spacetime of general relativity, the quantum fluctuations of spacetime in  
quantum gravity, and  the ideas of string/M-theory and loop quantum gravity that spacetime is an
approximation to something more fundamental.  
Changes in quantum theory have
reflected this evolution.  Non-relativistic quantum mechanics incorporates
Newtonian time in the Schr\"odinger equation and the second law of evolution.  Any one of the possible
timelike directions in Minkowski space can be used to describe the unitary
evolution of quantum fields and the results of different choices are
unitarily equivalent.  Quantum field theories in curved spacetimes based on
different foliations by spacelike surfaces are not generally unitarily
equivalent.  In quantum gravity there is no fixed spacetime through
which a state can unitarily evolve.  Quantum mechanics therefore needs to
be generalized for quantum gravity so that it does not require a fixed
spacetime foliable by spacelike surfaces. And, if spacetime is not fundamental, quantum mechanics will certainly need to be generalized for whatever replaces it. 

However, familiar quantum mechanics {\it also} needs to be generalized for cosmology. This generalization is needed so that quantum theory can apply to closed systems such as the universe as a whole containing both observers and observed, measuring apparatus and measured subsystems (if any).  These two generalizations can be connected in a common framework called {\it generalized quantum theory} which is abstracted from the consistent (or decoherent) histories formulation of the quantum mechanics of closed systems \cite{decohhist}.

The principles of generalized quantum mechanics were introduced in
Ref. \cite{Har91a} and developed more fully for example in
\cite{leshouches}.
The principles have been axiomatized in a rigorous
mathematical setting by Isham, Linden and others \cite{Isham}. Three elements are needed
to specify a generalized quantum theory:

\begin{enumerate}
\item The sets of {\it fine-grained histories}. These are the most
refined possible description of a closed system. 
\item The allowed {\it coarse grainings}. A coarse graining of a
set of histories is
generally a partition of that set into 
mutually exclusive classes $\{c_\alpha\}, (\alpha \ {\rm discrete)}$ called
{\it coarse-grained histories}.  The set of classes constitutes
a set of coarse-grained histories with each history labeled by the
discrete index $\alpha$. 
\item A {\it decoherence functional} defined for each allowed set of
coarse-grained histories which measures the interference between pairs of histories in the set and incorporates a theory of the initial
condition and dynamics of the closed system.  A decoherence
functional $D(\alpha^\prime,\alpha)$ must satisfy the following
properties.
\grouplabel{dfnl}
\begin{mathletters}
\begin{description}
\item{(i)} {\sl Hermiticity}: 
\begin{equation}
D(\alpha^\prime,\alpha)= D^*(\alpha,
\alpha^\prime)\ 
\end{equation}
\item{(ii)} {\sl Positivity}: 
\begin{equation}
D(\alpha, \alpha) \geq 0\ .
\end{equation}
\item{(iii)} {\sl Normalization}:
\begin{equation}
\sum\nolimits_{\alpha^\prime\alpha} D(\alpha^\prime, \alpha) = 1\ .
\end{equation}
\item{(iv)} {\sl The Principle of Superposition}:\hfill\break
If $\{\bar c_{\bar\alpha}\}$
 is a coarse graining of a set of histories
$\{c_\alpha\}$, that is, a further partition into classes $\{\bar
c_{\bar\alpha}\}$, then
\begin{equation}
D(\bar\alpha^\prime, \bar\alpha) =
\sum\limits_{\alpha^\prime\in\bar\alpha^\prime}
\ \sum\limits_{\alpha\in\bar\alpha} D(\alpha^\prime, \alpha)\ .
\end{equation}
\end{description}%
\end{mathletters}%
\end{enumerate}

Once these three elements are specified the process of prediction
proceeds as follows: A set of histories is said to (medium) decohere if
all the ``off-diagonal'' elements of $D(\alpha^\prime, \alpha)$ are
sufficiently small. The diagonal elements are the probabilities
$p(\alpha)$ of the individual histories in a decoherent set. These two
definitions are summarized in the one relation
\begin{equation}
D(\alpha^\prime, \alpha) \approx \delta_{\alpha^\prime\alpha}
p(\alpha)\ .
\label{basic} 
\end{equation}
As a consequence of (\ref{basic}) and properties (i)-(iv) above, the
numbers $p(\alpha)$ lie between zero and one, sum to one, and satisfy
the most general form of the probability sum rules
\begin{equation}
p(\bar\alpha) = \sum\limits_{\alpha\in\bar\alpha} p(\alpha) 
\end{equation}
for any coarse graining $\{\bar c_{\bar\alpha}\}$ of the set
$\{c_\alpha\}$. The $p(\alpha)$ are therefore probabilities. They are
the predictions of generalized quantum mechanics for the possible
coarse-grained histories of the closed system that arise from the 
theory of its initial
condition and dynamics incorporated in the construction of $D$.

Feynman's 1948 spacetime formulation of quantum mechanics \cite{1948}
supplies one route to constructing a fully four-dimensional generalized quantum theory of spacetime geometry.  The quantum mechanics of a non-relativistic particle moving in one dimension ($x=x(t)$) between time $t=0$ and time $t=T$  provides the simplest example. The particle's dynamics is assumed specified by an action functional $S[x(t)]$ and its initial quantum state is assumed to be a particular state vector $|\psi\rangle$. 

\begin{enumerate}

\item {\it Fine-grained histories:} These are all paths $x(t)$ between $t=0$ and $t=T$. 

\item{\it Coarse-grainings:} An allowed coarse graining is any partition of the paths into an exhaustive set of exclusive classes $c_\alpha$, ($\alpha$ discrete), each class being a {\it coarse-grained history}. For instance, the paths could be partitions by specifying a set of spatial intervals $\Delta_i$, $i= 1,2,\cdots$ and giving which two intervals $\alpha=(i,j)$ the particle passes through at two times. An example of a {\it spacetime coarse graining} is  provided by specifying a spacetime region $R$ and partitioning the paths into the class $c_0$ which never pass through $R$ and the class $c_1$ that pass through $R$ {\it some}time. 

\item{\it Decoherence functional:}  In a given set of coarse-grained histories $\{c_\alpha\}$ construct  {\it branch state vector} $|\psi_\alpha\rangle$ for each coarse grained history by summing $\exp(iS)$ over all the paths in $c_\alpha$ and applying that to the initial state $|\psi\rangle$, viz.
\grouplabel{nonrel}
\begin{mathletters}
\begin{equation}
|\psi_\alpha\rangle \equiv \int_{c_\alpha} \delta x \exp\{i S[x(t)]\hbar\} |\psi\rangle . 
\end{equation} 
The decoherence functional is 
\begin{equation}
D(\alpha',\alpha) = \langle\psi_{\alpha'}|\psi_\alpha\rangle .
\end{equation}
\end{mathletters}%
\end{enumerate}
 This spacetime formulation of non-relativistic quantum mechanics is easy 
to visualize, fully four-dimensional, manifests Lagrangian symmetries, and has
a close connection to the semiclassical approximation.  It incorporates both unitary
evolution and the reduction of the state vector in a unified way \cite{caves}.

The spacetime formulation is equivalent to usual Hamiltonian quantum
mechanics when the fine grained histories are {\it single valued in a time} as in non-relativistic quantum mechanics and  Minkowski space quantum field theory. This fully four-dimensional formulation generalizes usual quantum
mechanics when the histories do not have this property, for instance if there is no fixed time or the histories are not single valued in time. But in those cases we cannot expect to find a notion of state of the system at a moment of time or its unitary evolution through time. 

The talk illustrated these ideas with a series of model situations:

\begin{itemize}

\item Spacetime alternatives extended over time such as those defined by field averages over spacetime regions with  extent both in time and space \cite{stcg}.

\item Time-neutral quantum mechanics without a quantum mechanical arrow of time but with both initial and final conditions \cite{GH94}. 

\item Quantum field theory in fixed background spacetimes that are not foliable by spacelike surfaces such as spacetimes with closed timelike curves, spactimes exhibiting topology change, and evaporating black hole spacetimes \cite{nonchronal,evap}.

\item Histories that move backward in time such as those of a single relativistic particle moving in four-dimensional flat spacetime \cite{leshouches}.

\end{itemize}
For each of these examples the three ingredients of generalized quantum theory were exhibited --- fine grained histories, coarse graining, and decoherence functional. 

Building on the lessons of these examples, a  generalized quantum mechanics of quantum cosmological spacetime geometry can be sketched. The fine grained histories are closed four-dimensional cosmological metrics with four-dimensional matter field configurations upon them. The allowed coarse grainings are partitions of these histories into four-dimensional diffeomorphism invariant classes $c_\alpha$. A decoherence functional $D(\alpha',\alpha)$  is constructed using amplitudes defined by  sums over the histories in the classes $c_\alpha'$ and $c_\alpha$, initial and final wave functions of the universe, and an inner product linking amplitudes and wave functions. The semiclassical limit for geometry is provided by the steepest descents approximations to the sums over metrics.  What remains is a usual quantum field theory in the background spacetime described by the metric which gives the biggest contribution to these sums. Thus, familiar familiar quantum mechanics is recovered for those
initial conditions and those coarse-grainings in which spacetime is fixed, 
classical, and can supply the necessary time for unitary evolution.

A few points summarize the conclusion of the talk:

\begin{itemize}
\item Quantum mechanics can be generalized so that it is free from a fundamental notion of measurement, free of the need for a fixed background spacetime, and free from the `problem of time'.

\item General relativity as a theory of four-dimensional spacetime is more general than its 3+1 initial value problem. Simlarly, a fully four-dimensional formulation of quantum theory is more general than its 3+1 formulation in terms of states evolving unitarily through spacelike surfaces in a fixed background spacetime.

\item In a four-dimiensional generalized quantum mechanics of spacetime geometry there is no `problem of time', but there are also typically no states at a moment of time. 

\item In the context of a fully four-dimensional formulation of quantum theory, the  familiar  3+1 quantum mechanics of states evolving unitarily through spacelike surfaces is an approximation  that is appropriate for those initial conditions and those coarse grained descriptions in which spacetime geometry behaves classically and can supply  the  notion of time necessary to describe the evolution. 

\end{itemize}
\section*{Acknowledgment:} Preparation of this abstract was supported in part by the
National Science Foundation under grant PHY02-44764

\section*{References:}
These are largely references to the author's work on these subjects. References to the work of others can typically be found in them. Therefore this is not a bibliography of papers that address the questions of this paper, but rather pointers to references for the author's particular views.

\end{document}